# COSMOLOGICAL MODELS OF GAMMA-RAY BURSTS

CHARLES D. DERMER
*E. O. Hulburt Center for Space Research,*
*Naval Research Laboratory, Washington, D.C. 20375, U.S.A.*

and

THOMAS J. WEILER
*Department of Physics and Astronomy,*
*Vanderbilt University, Nashville TN 37235, U.S.A.*

**Abstract.** We review models of cosmological gamma-ray bursts (GRBs). The statistical and $\gamma$-ray transparency issues are summarized. Neutron-star and black-hole merger scenarios are described and estimates of merger rates are summarized. We review the simple fireball models for GRBs and the recent work on non-simple fireballs. Alternative cosmological models, including models where GRBs are analogs of active galactic nuclei and where they are produced by high-field, short period pulsars, are also mentioned. The value of neutrino astronomy to solve the GRB puzzle is briefly reviewed.

## 1. Statistical Premonitions

Soon after the first published report (Klebesadel *et al.*, 1973) of the discovery of gamma-ray bursts (GRBs), Usov and Chibisov (1975) discussed the statistical issues. The size distribution $N(>F)$ for GRBs produced by galactic disk neutron stars (NSs) bends from $N(>F) \propto F^{-1}$ for GRB fluences $F(\text{ergs cm}^{-2}) \ll F_b$ to $N(>F) \propto F^{-3/2}$ for $F \gg F_b$, with strong accompanying anisotropy of the GRB sky distribution towards the galactic disk when $F \ll F_b$. For burst energies $Q = 10^{38}Q_{38}$ ergs and scale heights of $300d_{21}$ pc, the break $F_b$ in the fluence size distribution occurs at $F_b \cong 10^{-5}Q_{38}d_{21}^{-2}$.

Two cosmological models were considered by Usov and Chibisov, and the point was made that a bend in the size distribution due to cosmological redshifting would occur for sources located at redshift $z \cong 1$, or source distances $d(\text{cm}) \cong c/H_0 \cong 10^7 d_{21} \equiv d_{28}$. For GRBs from supernovae with $Q_{38} = 10^{10}$, a flattened size distribution $N(>F) \propto k$, a constant, at low fluences followed solely from expansion effects. But with the bend in $N(>F)$ found at fluences $F < F_b \approx 10^{-9}$ ergs cm$^{-2}$, it would be quite difficult to detect due to the background galactic $\gamma$-ray glow. On the other hand, if $Q \sim 10^{52}$ ergs, then $F_b \approx 10^{-5}$ ergs cm$^{-2}$. The proposed model for this incredible energy release was the collapse of magnetic supermassive stars with masses $M \sim 10^5$-$10^6 M_\odot$, which was then being considered (Ozernoi and Usov, 1973; Prilutskii and Usov, 1975) as a model for the engine of active galactic nuclei (AGNs).

In 1983, van den Bergh published an early (the first?) Aitoff projection of the sky distribution of 46 GRBs. From their apparent isotropy, he con-cluded that the bursting sources were either non-exotic galactic black holes (BHs) or NSs with limiting detector sampling distances $d_d \approx 0.1 - 1$ kpc, or cosmological objects with $z \gtrsim 0.1$ in order to be in accord with the absence of GRB associations with M31, M33, SMC, LMC, Virgo, and other nearby galaxies. He therefore argued that $d_d \gtrsim 10^{27}$ cm for cosmological sources.

## 2. Fallout from the BATSE Discovery

The initial BATSE result (Meegan *et al.*, 1992) showed strong inhomogeneity to limiting peak count rates $C_p \cong 1$ ph cm$^{-2}$ s$^{-1}$ and limiting peak fluxes $\phi_p \cong 10^{-7}$ ergs cm$^{-2}$ s$^{-1}$ in the 50-300 keV band. In the discovery paper, $\langle V/V_{\max} \rangle = 0.348 \pm 0.024$. As reported by M. Briggs at this conference, the analysis of 657 GRBs gives $\langle V/V_{\max} \rangle = 0.330 \pm 0.011$. Briggs also reports that the dipole moment $\langle \cos \theta \rangle$ deviates less than $+0.9\sigma$ from 0, and the quadrupole moment $\langle \sin^2 b \rangle$ deviates by less than $-0.3\sigma$ from the value $1/3$. When the strong inhomogeneity of sources is coupled with burst directions in statistical accord with isotropy, galactic disk NS models are instantly demolished.

High-velocity NS models (e.g., Li and Dermer, 1992; Lamb, 1995; T. Bulik, this conference) remain an option, but require fine tuning of 3 parameters: (i) the sampling distance or intrinsic GRB luminosity; (ii) a delayed turn-on parameter; and (iii) strong suppression of the bursting rates of slow ($v \lesssim 800$ km s$^{-1}$) NSs. Tuning of parameter (iii) is particularly improbable, so that the statistical evidence for isotropy favors the cosmological model by wide margins.

## 3. Statistics Post-BATSE

But do the simplest cosmological models fit the observed size distribution of GRBs? In terms of the peak flux distribution in the energy range 100-500 keV (see Fenimore *et al.*, 1993a), $N(> \phi_p) \propto \phi_p^{-0.8}$ for $10^{-7} \lesssim \phi_p$ (ergs cm$^{-2}$ s$^{-1}$) $\lesssim 2 \times 10^{-6}$, and $N(> \phi_p) \propto \phi_p^{-1.5(\pm 0.1)}$ at larger values of $\phi_p$ when combined with the PVO data. The earliest analyses of the statistics of GRBs after the announcement of the original BATSE results assumed no source evolution with $z$ for either luminosity or comoving number density, monoluminous GRBs, power-law or broken power-law spectra, and unbeamed sources that uniformly emit in all directions. Good fits to the BATSE data were in all cases possible, with model fits implying limiting redshifts $z_B$ of the faintest detectable BATSE GRBs, peak GRB luminosities $L_p$ and bursting rates $\rho$. Table 1 gives the model results for the early analyses. The main difference between these models is in the implied values of $L_p$, which depends on the assumed bandwidth. The high value of Dermer (1992) is due to the use of a broken power-law spectrum and a GRB bandwidth extending to 100 MeV, whereas the other analyses restrict $L_p$ to the BATSE bandpass.

Wickramasinghe *et al.* (1993) considered different cosmological models with $\Omega = 0.1$ and $\Omega = 1.0$, but could not discriminate between the two cosmologies from the BATSE GRB data. Tamblyn and Melia (1993) performed a K-correction for different GRB detectors, their associated bandwidths, limiting thresholds, and trigger criteria. A broken power law GRB spectrum was used to approximate the curved GRB spectra, and reconcile the detection rates by PVO, SMM, KONUS, SIGNE, APEX, and BATSE.

TABLE I

Inferred Properties of Mono-luminous, Non-Evolving, Unbeamed GRBs.

| Reference | $z_B$ | $L_p$ (ergs s$^{-1}$) | Bursting rate $\rho$ |
|---|---|---|---|
| Piran (1992) | 1 | $10^{50}$ (1 s GRB) | $3 \times 10^{-7}$ yr$^{-1}$ galaxy$^{-1}$ |
| Mao and Paczyński (1992) | 1.5 | $2 \times 10^{51}$ | $2 \times 10^{-6}$ yr$^{-1}$ ($L_*$ galaxy)$^{-1}$ |
| Dermer (1992) | 1.2 | $4 \times 10^{51}$ | $10^{-7}$ yr$^{-1}$ Mpc$^{-3}$ |
| Fenimore *et al.* (1993a) | 0.8 | $5 \times 10^{50}$ | $2.4 \times 10^{-8}$ yr$^{-1}$ Mpc$^{-3}$ |

Burst luminosity functions provide too much parameter freedom, and are therefore not well constrained. The only evident restriction is that the luminosity function be not sufficiently broad and gently varying so as to erase the rather abrupt change in slope in the observed size distribution from a $-0.8$ slope to a $-1.5$ slope. Beaming effects have been considered by, for example, Yi (1993) and Yi and Mao (1994). The flux density $S$ (ergs cm$^{-2}$ s$^{-1}$ MeV$^{-1}$) observed from a plasma blob travelling with Lorentz factor $\Gamma$ and speed $\beta c = (1 - \Gamma^{-2})^{-1/2} c$ with respect to the stationary frame in the Hubble flow goes as $S \propto L_{com} \mathcal{D}^{3+\alpha}$. Here it is assumed that the source isotropically radiates a spectrum with energy spectral index $\alpha$ and total photon luminosity $L_{com}$ in the comoving frame. The Doppler factor $\mathcal{D} = [\Gamma(1 - \beta\mu)]^{-1}$, where $\theta = \arccos \mu$ is the angle between the jet axis and the line-of-sight to the observer. Values of $\langle V/V_{\max} \rangle \approx 0.33$ are found when $L_{com} \cong 10^{51}/\Gamma^{3+\alpha}$ ergs s$^{-1}$, which corresponds to fairly weak luminosities indeed if $\Gamma \sim 10^2 - 10^3$. The total number of sources must be increased by a factor $\sim 4\Gamma^2$ to take into account those which are not directed into our viewing direction.

## 4. Getting the $\gamma$ Rays Out: The Need for Speed

Table 1 shows that at distances $d = 10^{28} d_{28}$ cm, GRBs must have source luminosities

$$L(\text{ergs s}^{-1}) = d^2\phi\delta\Omega \approx 10^{50}\zeta d_{28}^2\phi_{-6}\delta\Omega \ , \tag{1}$$

where $\zeta$ is a bandwidth correction factor, $\phi = 10^{-6}\phi_{-6}$ ergs cm$^{-2}$ s$^{-1}$ is the measured energy flux, and $\delta\Omega$ is the solid angle into which the emission is beamed. From the Elliot-Shapiro (1974) relation, updated for Klein-Nishina corrections on the radiation force (Dermer and Gehrels, 1995), this immediately rules out unbeamed Eddington-limited accretion models by $\sim 9$ orders of magnitude for a variability time scale $\delta t_v = 10$ ms.

Gamma rays can only escape from the source region if its pair production optical depth $\tau_{\gamma\gamma}$ through the reaction $\gamma\gamma \to e^+e^-$ is $\ll 1$. For MeV $\gamma$ rays, this is essentially equivalent to requiring that the photon compactness for unbeamed sources

$$\ell = \frac{L}{R}\frac{\sigma_T}{m_ec^3} \approx \frac{10^{12}\zeta d_{28}^2\phi_{-6}}{\delta t_v(s)} \lesssim 1 \ ; \tag{2}$$

hence the formation of a pair fireball is inevitable for stationary sources at cosmological distances (Piran and Shemi, 1993).

Bulk relativistic motion relieves $\gamma$-ray opaqueness in two ways (Krolik and Pier, 1991). For plasma blobs radiating isotropically in the comoving frame, the comoving frame photon energy $\epsilon_{\text{com}}(= h\nu_{\text{com}}/m_ec^2)$ is reduced by a factor $\Gamma$ from the observed energy $\epsilon_{\text{obs}}$, and the inferred photon energy densities in the comoving frame are reduced by a factor $\Gamma^{3+\alpha}$ compared to inferences from stationary sources. To avoid $\gamma\gamma$ attenuation, it is sufficient to require that $\epsilon_{\text{com}} < 1$ to avoid the pair production threshold, so that one requires that $\Gamma >$ the maximum observed photon energy $\epsilon_{\text{obs,max}}$. In view of high energy photons observed with SMM and COMPTEL, and even $> $ GeV photons with EGRET (e.g., Hurley *et al.* 1994), it seems more reasonable to consider power-law photon spectra extending without high-energy limits in the source. Baring (1993) shows that for selected GRBs, the reduction in the comoving frame energy density implies that $\Gamma$ must be $\gtrsim 10^2$ to avoid $\gamma\gamma$ absorption.

One must also make sure that photons emitted at earlier times do not constitute an additional source of opacity. Fenimore *et al.* (1993b) and Woods and Loeb (1995) have considered an expanding shell geometry and find again that values of $\Gamma \sim 10^2$ or greater are again needed to get the $\gamma$ rays out. Everything points to expansion at relativistic speed.

## 5. Simple Fireballs

Cavallo and Rees had already treated in 1978 a system which naturally produced plasma expanding relativistically outwards. The most important processes in their cosmic fireball were pair production and annihilation and Compton scattering. They identified an important quantity that related the total fireball energy $Q_f$ to the baryon mass $M$ through the expression

$$Q_f = \eta M c^2 \; . \tag{3}$$

The term $\eta$ is essentially the entropy or photon energy per baryon, or the baryon-loading parameter. Because the important parameters in the fireball were a baryon or proton optical depth $\tau_p = n_p \sigma_T R_0$, and $\tau_{\gamma\gamma}$ or the fireball compactness $\ell \propto L/R_0 c$, they identified the four fireball types shown in Table 2. The term $R_0$ refers to the radius of the region into which $Q_f$ is injected, and $n_p$ is the mean proton number density.

TABLE II
Simple Fireball Types (Cavallo and Rees, 1978).

| $\tau_{\gamma\gamma} \gg 1$, $\tau_p \gg 1$ | $\tau_{\gamma\gamma} \gg 1$, $\tau_p \ll 1$ |
|---|---|
| Baryon-Dominated Fireball | Pure Fireball |
| $\tau_{\gamma\gamma} \ll 1$, $\tau_p \gg 1$ | $\tau_{\gamma\gamma} \ll 1$, $\tau_p \ll 1$ |
| Thermalized Radiation | Line-Production Region |

For the baryon-dominated fireball, the terminal speeds $\beta_t c$ or terminal Lorentz factors $\Gamma_t$ of an expanding fireball were found to be

$$\beta_t \cong \eta^{1/2} \; , \; \eta \ll 1 \tag{4a}$$

$$\Gamma_t \cong \eta \; \; , \; \eta \gg 1 \; . \tag{4b}$$

Shemi and Piran (1990) introduced the radius $R_{\text{thin}}$ where the fireball becomes optically thin, so that Eq. (4b) is generalized to

$$\Gamma_t = \frac{1+\eta}{1 + \frac{R_0}{R_{\text{thin}}}\eta} \rightarrow \begin{cases} R_{\text{thin}}/R_0, & \text{if } \eta \gg R_{\text{thin}}/R_0 \\ 1+\eta, & \text{if } \eta \ll R_{\text{thin}}/R_0 \end{cases} \; . \tag{5}$$

Although Cavallo and Rees had made the connection of fireballs to GRBs, it was Paczyński (1986) who pointed out two remarkable coincidences between fireballs and GRBs. The first was that if some small fraction $\approx 0.1\% - 1\%$ of the $10^{53} - 10^{54}$ ergs available as the rest mass energy of a NS were liberated as $\gamma$ rays from cosmological distances, then a fluence $F \approx 10^{-6} Q_{51} d_{28}^{-2}$ ergs cm$^{-2}$ would be measured near Earth. NS phase transitions and coalescing NSs were proposed as possible burst origins. The second coincidence was that if this energy were injected into a source size $R \approx 10$ km on a time scale $\delta t_i$ of a second or less, then the blackbody temperature

$$T_0(\text{MeV}) \approx 3 \frac{(Q_{51}/\delta t_i)^{1/4}}{(R_0/10\text{km})^{1/2}} , \qquad (6)$$

nicely in the $\gamma$-ray range.

Paczyński solved the steady-state spherically symmetric flow equations for continuous injection of energy with no baryon loading. For adiabatic expansion, $\Gamma(R) \cong R/R_0$, and the comoving frame temperature $T_{\text{com}}(R) \cong T_0 R_0/R$. An observer would see Doppler-shifted emission with effective temperature $T_{\text{obs}} \cong \Gamma(R)T_{\text{com}}(R) \approx T_0$, so that the emission from the expanding fireball would also peak at $\gamma$-ray energies. Paczyński recognized that a thermal spectrum does not give good fits to GRB spectra, and suggested that a cool envelope would scatter $\gamma$ rays to lower energies to produce a photon flux $\phi(\epsilon) \gtrsim \epsilon^{-1}$, as observed.

Goodman (1986) considered a more realistic time-dependent bursting source. The calculated spectrum is slightly broader than a Planckian. He also pointed out that the minimum variability time scale would be $\approx \Delta R/c \approx R_0/c$ and that the GRB duration would be $\approx R_{\text{thin}}/\Gamma_i^2 c$, which could in principle be as short as $R_0/c$.

## 6. Merger Rates and Scenarios

The discovery of the binary radio pulsar system PSR 1913+16 (Hulse and Taylor, 1975) with a merger time scale $t_{\text{merge}} \approx 3 \times 10^8$ yr $\ll$ the Hubble time $t_{\text{H}} \approx 10^{10}$ yrs, prompted Clark et al. (1979) to make a first estimate of NS-NS merging rates. Because there was one such system for the 300 then-known isolated radio pulsars, and because the NS birthrate $\dot{N}_{\text{NS}} \approx 0.02$ yr$^{-1}$, this implies a birthrate of binary NSs equal to $\dot{N}_{\text{NS-NS}} \approx 10^{-4}$ yr$^{-1}$. This is also the merging rate in steady-state if $t_{\text{merge}} \ll t_{\text{H}}$.

Narayan et al. (1991) performed a better treatment of selection biases in binary and isolated pulsar searches, and also considered the additional information provided by 3 new binary pulsars. They estimated $\dot{N}_{\text{NS-NS}} \approx 10^{-5} h_0(\text{kpc})$ yr$^{-1}$, where the unknown binary pulsar scale height $h_0$ is estimated at a few kpc. From the observed massive X-ray BH binary systems such as Cyg X-1 and LMC X-3, they also estimated the birthrate for NS-BH binaries to be $\dot{N}_{\text{NS-BH}} \approx 10^{-4.5}$ yr$^{-1}$. For 10% of the binary NS sources having $t_{\text{merge}} < t_{\text{H}}$, this implies $\dot{N}_{\text{NS-NS}}^{\text{merge}} \sim 10^{-6}$ yr$^{-1}$, which is consistent with the implied GRB burst rates in Table 1 provided that the sources are not strongly beamed.

Tutukov and Yungelson (1993) estimate higher merger rates than Narayan et al. (1991). For our galaxy, they find that $\dot{N}_{\text{NS-NS}}^{\text{merge}} \approx 3 \times 10^{-4}$ yr$^{-1}$ and $\dot{N}_{\text{NS-BH}}^{\text{merge}} \approx 10^{-5}$ yr$^{-1}$. This shows that order-of-magnitude uncertainty still remains. New estimates have recently been given by Lipunov *et al.* (1995) and K. Postnov at this conference.

The sequence of events accompanying the merging of two NSs was considered by Eichler *et al.* (1989), Haensel *et al.* (1991), and Narayan *et al.* (1992). Besides being a model for GRBs (Paczyński's earlier suggestion), Eichler *et al.* speculated that merging NSs would be sources of r-process elements, neutrino ($\nu$) bursts and gravitational waves. The generation of the fireball is a consequence of prolific $\nu$ production in binary coalescence through the reaction $\nu\bar\nu \rightarrow e^+e^- \leftrightarrow 2\gamma$, as had been highlighted earlier for collapse events by Goodman *et al.* (1987). The $\nu$ and $\bar\nu$ originate from pp and pn bremsstrahlung during the contorting tidal heating events just preceding binary coalescence – precisely the same reactions which so quickly cool a NS just after birth.

The efficiency of this reaction to produce pairs is estimated at

$$\xi \approx f\sigma_{\bar\nu\nu}\frac{Q_\nu}{\bar E_\nu}\frac{1}{R_0 c\delta t_i} \sim 10^{-3} \tag{7}$$

(Piran *et al.*, 1992), where $Q_\nu$ is the total energy in $\nu$ emitted with average energy $\bar E_\nu$ in a region with size scale $R_0$ during injection timescale $\delta t_i$, $f$ is a geometrical factor, and $\sigma_{\bar\nu\nu} \sim 10^{-44}(\bar E/10~{\rm MeV})^2~{\rm cm}^2$ is the cross section for the reaction $\nu\bar\nu \rightarrow e^+e^-$.

Of the $10^{53}$-$10^{54}$ ergs of energy released in binary NS coalescence, some $2 \times 10^{53}$ ergs appear in the form of $\nu$ and $\approx 8 \times 10^{52}$ ergs in the form of gravitational waves (Harding, 1994; see also Narayan *et al.*, 1992). With $\xi \sim 10^{-3}$, this implies that $\approx 10^{50}$ ergs is injected as fireball energy on the very short ($\sim$ ms) event during which the final coalescence occurs and the bulk of the energy transfer is accomplished, in large measure of course by gravitational wave damping. Numerical simulations (M. Ruffert, this conference), however, are finding smaller conversion efficiencies. If these low efficiencies are correct, then beaming of the pair fireball may be required to produce the observed $\gamma$ bursts, with the attendant required increase in $\rho$.

## 7. Not-so-simple Fireballs

During the coalescence event, tidal heating not only generates a huge luminosity in $\nu$, but also a photon-driven mass loss when the radiation pressure ejects a wind. The tidal heating rate

$$\dot E_t = \frac{GM^2}{R_*}(\frac{R_*}{D})^6\frac{1}{qt_0} \propto 10^{56}D_6^{-15/2}~{\rm ergs~s}^{-1} \tag{8}$$

(Mészáros and Rees, 1992a), where $D = 10^6 D_6$ cm is the separation radius, $R_*$ is the stellar radius, $t_0$ is the orbital time scale, and $q$ is a correction factor. An ejected mass $\Delta M \simeq 10^{-3}M_\odot$ is calculated. Only $\gtrsim$ few$\times 10^{-9}M_\odot$ is, however, sufficient to impede $\gamma$-ray production, because then the bulk of the fireball energy goes into the kinetic energy of the baryons (Shemi and Piran, 1990) and the burst becomes radiatively inefficient.

Baryon contamination can be avoided by anisotropic baryon expulsion. Mészáros and Rees (1992b) suggested that a baryon-free zone would exist above and below the contact point of two coalescing NSs, and in this region the $\nu\bar{\nu}$ production would eject a pair jet. They also suggested that gravitational bending of $\nu$ and $\bar{\nu}$ trajectories would produce a baryon-free zone in the side opposite the NS during a NS-BH merger. Mochkovitch et al. (1993) argue that a NS would be completely deformed into a torus during its merger with a BH, so that a natural axis of symmetry is formed along which $\nu\bar{\nu}$ annihilation occurs and a relativistically expanding $e^+e^-$ wind is expelled. More detailed calculations by Mochkovitch et al. (1995) for prescribed geometries of the $\nu$-production region showed that the pair luminosity per unit solid angle $L_{\nu\bar{\nu}\to e^+e^-} \sim 10^{50}$ ergs sr$^{-1}$. The associated mass loss for different geometries could produce a situation where $\eta$ decreases or the baryon loading increases with increasing $\nu$ temperature in the $\nu$ production region.

More detailed studies of fireball evolution have been treated both analytically (Piran, 1994; Katz, 1994a) and numerically (Mészáros et al., 1993). The important point is that the fireball goes through both a radiation- and a matter-dominated phase, as in the early universe. In its late evolution it also goes through a SN-like deceleration phase when the fireball sweeps up a sufficient quantity of matter. The fireball deceleration radius occurs at

$$R_d \simeq \left(\frac{3Q}{4\pi n m_p c^2 \Gamma^2}\right)^{1/3} \cong 5 \times 10^{15} Q_{51}^{1/3} n^{-1/3} \Gamma_3^{-2/3} \text{ cm}, \qquad (9)$$

(Rees and Mészáros, 1992; Mészáros and Rees, 1993a), with a deceleration time scale

$$t_d \simeq 0.1 Q_{51}^{1/3} n^{-1/3} \Gamma_3^{-8/3} \text{ s}. \qquad (10)$$

Here $\Gamma_t = 10^3 \Gamma_3$ is the terminal Lorentz factor of the fireball before entering the deceleration phase, and $n$ is the density of the surrounding interstellar medium. Some 80% of the mass is concentrated in a thin shell whose lab frame width remains constant until the shell becomes transparent. A reverse shock into the fireball accompanies the blast wave as it moves into the ISM. The reconversion of the kinetic energy of the baryons into radiation after interaction with the external medium helps relieve the problem of baryon contamination.

Spectral modelling results (Mészáros and Rees, 1993b) for blast wave models of GRBs so far rely on synchrotron and synchrotron self-Compton processes, and the results are therefore very sensitive to the assumed mag-netic field $B$. Mészáros *et al.* (1994) consider spectral production in three scenarios. For the frozen-in field model, the ejecta is assumed to have an entrained $B$-field which is some fraction of equipartition with the total energy density of the fireball. Dynamo amplification of $B$ is produced by a reverse shock. For the turbulent model, turbulent amplification of the $B$ is produced both in the blast wave as well as by the reverse shock in the ejecta, but only in the blast wave for the piston model. Model results show extremely weak emission at all energies below the $\gamma$-ray regime, but no explanation is given concerning the rough uniformity of the $\nu F_\nu$ peaks in GRBs near 1 MeV. Brainerd (1994) attributes this uniformity to Compton scattering by a surrounding optically thick medium with column density $N_H \sim 10^{25}$ cm$^{-2}$, which might occur if the GRB sources are buried in molecular clouds in the cores of galaxies.

The fine scale structure in GRB spectra could be due to Rayleigh-Taylor convective instabilities in the expanding blast wave (Waxman and Piran, 1994), overtaking shocks from unsteady outflow during the event triggering a GRB (Rees and Mészáros, 1994), or even to patchy structure of the surrounding ISM. The delayed GeV emission observed in the 17 Feb 1994 GRB (Hurley *et al.*, 1994) has led to a rethinking concerning the contribution of different phases of fireball evolution to the observed GRB time history (Mészáros and Rees 1994). Perhaps the main MeV portion of the burst is the accelerating phase, whereas the long-duration high energy tail occurs when the blast wave is decelerated by the surrounding medium. Katz (1994b) suggests that the delayed high-energy radiation is due to the interaction of the blast wave with a dense cloud and the production of $\gamma$ rays through secondary production reactions, notably $pp \rightarrow \pi^o \rightarrow 2\gamma$.

## 8. Other Models

Models where GRBs are analogs of AGNs (McBreen *et al.*, 1993; Dermer and Schlickeiser, 1994; Shaviv and Dar, 1995) are favored in view of the similarities between the two cases, such as the peaking of the $\nu F_\nu$ spectrum in the $\gamma$-ray regime, the rapidly varying $\gamma$-ray emission, and the strong flaring behavior of blazar AGNs. Roland *et al.* (1994) have developed this model most fully and calculate complex time profiles due to the perturbation of the jet by a Solar mass NS or BH accreting onto a $10^3 - 10^6 M_\odot$ BH. The differences and problems may, however, be greater than the similarities.

The $\nu F_\nu$ spectral peaks for AGNs ranges over at least three decades in energy compared to only $\sim 1$ decade for GRBs. The compactnesses differ by $\sim 8$ orders of magnitude, even for the most luminous and rapidly variable blazars such as PKS 0528+134. A catastrophic event, such as tidal disruption of a star (Carter, 1992), could produce the flaring GRB behavior from a massive accreting black hole. But except for some carefully contrived geo-metrical constructions, the minimum variability time scale corresponds to the dynamical time scale associated with the Schwarzschild radius of the BH. And if a BH with mass $\lesssim 10^3 M_\odot$ is needed to agree with millisecond variability observed in GRBs, then this model reduces to the NS-BH merger scenario. Also, no AGNs have been found in GRB error boxes.

Usov (1992; 1994) has proposed a strong field millisecond pulsar model for GRBs. During the collapse of a white dwarf to a NS, it is argued that $B \rightarrow 10^{15}$ G from flux freezing and the period $P \rightarrow 1$ ms from angular momentum conservation. For such a system, the magnetic dipole luminosity can exceed $10^{51}$ ergs s$^{-1}$ and the gravitational quadrupole luminosity can exceed $10^{55}$ ergs s$^{-1}$. Because the energy release is so rapid, a pair fireball is formed, leading to a situation essentially equivalent to the NS merger scenarios. Collapse events leading to a pair or a Poynting-dominated MHD wind have also been considered by Woosley (1993) and Thompson (1994).

## 9. Neutrinos from Burst Hell

In GRB models with pion-decay (e.g., Paczyński & Xu, 1994), the $\nu$ energies $E_\nu$ range from an MeV to at least a few GeV, with flavor content in the ratio $\nu_e : \nu_\mu : \nu_\tau = 1 : 2 : 0$. In models with emission from superconducting cosmic strings (e.g. Plaga, 1994), $\nu_e : \nu_\mu : \nu_\tau = 1 : 1 : 1$, with $E_\nu$ up to $\sim 10$ TeV. Observation of $\nu$-flavor ratios could discriminate between these models. A measurement of the $\nu$ to $\gamma$ flux ratio ($\equiv \chi$) also provides important information about the bursting sources. Depending on optical depths, the $\nu$ luminosity can greatly exceed that in $\gamma$ rays ($\chi \gg 1$ is possible for hadronic models of AGNs). On the other hand, if the $\gamma$ rays have a purely electromagnetic origin, $\chi$ would be near zero. Weak experimental limits from underground experiments already exist for the $\nu$ flux associated with GRBs (Miller et al., 1994; Fukuda et al., 1994; Becker-Szendy et al., 1995), which translate into a limit on $\chi \lesssim$ a few $\times 10^3$ (the precise limit is spectrum-dependent).

Next year, the Super Kamiokande detector will improve the sensitivity for events below a few GeV by $\sim 10$. The effective volumes of the Baikal, AMANDA, NESTOR, and DUMAND ice/water instruments now under construction will be $\approx 25$ to 100 times that of the underground instruments, if the GRB $\nu$ spectrum extends to $E_{\nu,max} \gtrsim$ tens of GeV. The estimated $\nu$ counting rate for each detector is $\sim 10^{-4} \chi$ per burst, assuming an $E^{-2}$ spectral shape, 1 photon ($> 1$ MeV) per cm$^2$ per GRB (Fishman et al. 1994), and a photon spectrum reaching $E_{max} \sim 1$ TeV. With a GRB detection rate of $\approx 1$ per day, the expected number of correlated, detected $\nu_\mu$s per year is $\sim 10^{-2} \chi$. Input assumptions are imprecise, so this rate could either be easily detectable or beyond experimental reach.

If the energy threshold for the neutrino telescope is $\ll 1$ TeV, then the expected number of $\nu$ counts is $\sim 5 \times 10^{-3} \chi$ per GRB or $\approx 2\chi$ per year.

A possible high rate of multiple events from a single GRB is also possible: the nearest 0.1% of GRBs might produce the spectacular signature of 50 muons once every two years if $\chi \approx 100$. With these rates, the nature of the source and properties of neutrinos (masses, lifetimes, charge, speed) could be determined, and the weak equivalence principle could be tested by comparing Shapiro delays of photons, neutrinos, and anti-neutrinos passing the nucleus of our Galaxy. The large distance scale of GRBs allows studies of long oscillation lengths, with possible determination of tiny $\nu$ masses. Because the oscillation phase from $z \gtrsim 0.5$ sources is sensitive to the cosmological model, $\nu$ detection from GRBs could also test standard cosmology and further constrain $H_0$ and $\Omega_0$ (see Weiler *et al.*, 1995 for details).

## 10. Finally

The coalescing NS model remains the favorite cosmological model for GRBs because the sources are known to exist, the right amount of energy is involved, and the expected merger rates are in agreement with the required GRB rates. Absence of recurrence is evidently in accord with the data, although lensing events and triple star systems could provide escape hatches should GRB recurrence be demonstrated. Neutrino telescopes can probe the heart of the explosion and discriminate between models. But only the identification of counterparts at other wavelength ranges will conclusively solve the burst puzzle.